# A Unified Theory of Free Energy Functionals and Applications to Diffusion

Andrew B. Li[1], Leonid Miroshnik[2], Brian D. Rummel[2], Ganesh Balakrishnan[3], Sang M. Han[2,3], and Talid Sinno[4, a)]

[1]Physics and Astronomy, University of Pennsylvania

Philadelphia, Pennsylvania 19146, USA

[2]Chemical & Biological Engineering, University of New Mexico

Albuquerque, New Mexico, 87131, USA

[3]Center for High Technology Materials, University of New Mexico

Albuquerque, New Mexico, 87131, USA

[4]Chemical & Biomolecular Engineering, University of Pennsylvania

Philadelphia, Pennsylvania 19146, USA

a) Electronic mail: talid@seas.upenn.edu

**Abstract**

Free energy functionals of Ginzburg-Landau type lie at the heart of a broad class of continuum dynamical models, such as the Cahn-Hilliard and Swift-Hohenberg equations. Despite the wide use of such models, the assumptions embodied in the free energy functionals are frequently either poorly justified or lead to physically opaque parameters. Here, we introduce a mathematically rigorous pathway for constructing free energy functionals that generalizes beyond the constraints of Ginzburg-Landau gradient expansions. We show that the new formalism unifies existing free energetic descriptions under a single umbrella by establishing the criteria under which the generalized free energy reduces to gradient-based representations. Consequently, we derive a precise physical interpretation of the gradient energy parameter in the Cahn-Hilliard model as the product of an interaction length scale and the free energy curvature. The practical impact of our approach is demonstrated using both a model free energy function and the silicon-germanium alloy system.

## Introduction

The principal goal of classical field theories, such as Ginzburg-Landau (GL) type[1] and classical density functional theory (DFT)[2], is to mathematically describe a system's free energy in terms of some order parameter(s) and consequently drive a continuum dynamical model (e.g., Cahn-Hilliard (CH) equation, phase field[3–5]). These continuum models play a central role in our understanding and mathematical modeling of the natural world in a vast range of applications spanning nucleation[6], dendritic growth[7], self-assembly[8], intracellular organization[9,10], and brain cortex dynamics[11]; see Fig. 1. Moreover, they have become objects studied in their own right as distinct classes of stochastic PDEs[12]. Even in situations where the underlying microscopic physics may be described explicitly at the atomistic scale (e.g., molecular dynamics[13], Langevin dynamics[14], or Glauber dynamics[15] driven by interatomic potentials), the hydrodynamic/probabilistic limits of these descriptions are often described in terms of free energy gradient flows[16]. Consequently, constructing a free energetic description within a unified and physically comprehensive framework is a centrally important task for the continuum modeling of dynamical systems.

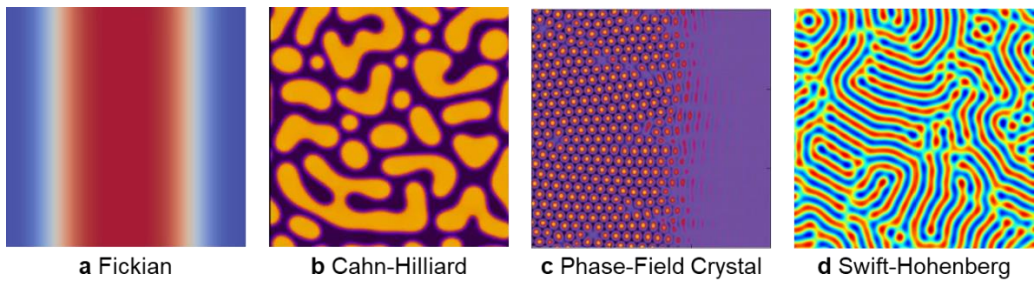

**Fig. 1 | Representative pattern structures predicted by various free energy functional-based continuum models.** **a** Fickian diffusion[17], **b** Cahn-Hilliard equation[18], **c** Phase-field crystal[19], and **d** Swift-Hohenberg equation[20].

The broad success of GL modeling notwithstanding, physical interpretations of GL free energy parameters are variably ambiguous except in a few idealized cases[21,22]. This difficulty arises principally from the phenomenological supposition that the free energy is expressible in terms of a sequence of gradients of one or more order parameters[23]. Some insightful attempts have been made to derive GL free energies with a more explicit physical basis, most notably classical DFT[24], which relies on a liquid reference state[25]. One example is the Giacomin-Lebowitz model of phase segregation[15], which has gained much attention in recent years as a non-local GL type theory that is physically interpretable. In another instance, a simplified classical DFT formulation, which leads to Swift-Hohenberg free energies, has given rise to the popular phase-field crystal (PFC) approach[25,26]. The PFC framework has been proposed as a bridge[27] between classical dynamical DFT and phase-field models, although the numerous simplifications embodied within it have been observed to lead to various unphysical predictions[25].

Here, we propose a generalization of GL type theory that addresses the challenges discussed above. We show that the new formalism relaxes the locality assumption in GL theory by removing the constraint that the free energy be strictly defined in terms of gradients. We also demonstrate, using specific examples, how the generalization reduces to widely employed models, such as the Cahn-Hilliard free energy, and in so doing obtain explicit criteria for their validity. Perhaps most practically, we also show that the generalized approach naturally leads to physically interpretable parameters while at the same time retaining the inherent multiresolution nature of the GL type framework. In this paper, we limit our analysis to species diffusion (i.e., conserved gradient flow) to demonstrate these features but emphasize that the free energy construction itself is entirely general.

## Continuum Modeling of Diffusion

The standard continuum diffusion equation for species $i$ is given by

$$\frac{\partial c_i}{\partial t} = \sum_j \nabla \cdot \left( \mathbf{M}_{ij} \cdot \nabla \beta \mu_j \right) + \epsilon = \sum_j \nabla \cdot \left( \mathbf{M}_{ij} \cdot \nabla \beta \frac{\delta F}{\delta c_j} \right) + \epsilon, \quad (1)$$

where $c_i$ is the concentration, $\mathbf{M}_{ij}$ is the mobility matrix, $\nabla \mu_j$ is the driving force due to a generalized chemical potential $\mu_i$, and $\epsilon$ is a thermal noise term that satisfies the fluctuation-dissipation theorem[28]. The generalized chemical potential, $\mu_i \equiv \frac{\delta F}{\delta c_i}$, is defined as the variational derivative of the free energy of the system, $F$. In the present analysis, we neglect the noise term and focus on deterministic evolution. For isotropic single-component diffusion, $\mathbf{M} = cD\mathbf{I}$, where $D$ is the self-diffusivity. Similarly, for isotropic binary interdiffusion, $\mathbf{M} = V_{atom}^2 (c_A^2 D_A + c_B^2 D_B)\mathbf{I}$ in the lattice reference frame, where the subscripts indicate atomic species, and $V_{atom}$ is the molar/atomic volume. Although here we only explicitly consider single and binary component cases and drop the indices $i$ and $j$, an extension to multicomponent cases[29] is straightforward.

The free energy[1], $F[c] \equiv \int f([c], \mathbf{r}) dV$, is most generally assumed to be a functional of the composition/density profile $[c]$, where $f([c], \mathbf{r})$ is the position-dependent free energy density functional. Without loss of generality, $F$ can be decomposed into ideal and excess contributions, i.e.,

$$F = F^{id} + F^{ex} = \int f^{id} dV + \int f^{ex} dV, \quad (2)$$

where $f^{id}$ and $f^{ex}$ are the corresponding free energy densities. We do not consider external fields explicitly as they contribute one-body terms that, along with $f^{id}$, do not modify the theory. Note that the separation of $F$ into ideal and excess components is natural as they arise from different aspects of the Brownian motion that generates the diffusion equation. For example, classical DFT relates $f^{ex}$ to the Ornstein-Zernike relation/direct correlation[30] function using the liquid/homogeneous state as a reference. GL type theories do not usually consider this separation explicitly and assume $F$ can be expanded directly with respect to gradient terms[31]. Below we present an alternative framework for constructing the functional $f^{ex}$ in terms of a sequence of convolution kernels that are directly linkable to the microscopic physics. Importantly, this framework requires no inherent assumptions or constraints be placed on $F[c]$ and can be linked formally to both GL type and classical DFT theories.

**General Free Energy Functional**

Consider a discretized compositional profile where $f^{ex}$ is to be evaluated at position $\mathbf{r}$, and $\{c_i\} \equiv \{c(\mathbf{r} + \Delta \mathbf{r}_i)\}$ is the set of compositions that are $\{\Delta \mathbf{r}_i\}$ away from $\mathbf{r}$. For the special case where the discretization corresponds to a crystal lattice, $c_i$ denotes the probabilities of an atomic site being occupied by an atom. Assuming that $f^{ex}([c], \mathbf{r})$ is analytical with respect to variations in $[c]$, it is possible to carry out a Taylor expansion with respect to any reference compositional profile. Specifically, we seek an expression that relates the functional $f^{ex}([c], \mathbf{r})$ to the function $\hat{f}^{ex}(c)$—the evaluation of $f^{ex}([c], \mathbf{r})$ at constant composition $c$—whose information can be obtained from equilibrium thermodynamic state variables and phase diagrams. $f^{ex}([c], \mathbf{r})$ is then given by (see Methods A)

$$f^{ex}([c], \mathbf{r}) = \hat{f}^{ex}(c) + \sum_m \frac{1}{m!} \int \rho_m^c \prod_{i=1}^m (c_i - c) dV_i, \qquad (3)$$

where $c = c(\mathbf{r})$, $c_i = c(\mathbf{r} + \Delta \mathbf{r}_i)$, and the $\rho_m^c(c, \{\Delta \mathbf{r}_i\})$ are constrained by

$$\left. \frac{\partial^m \hat{f}^{ex}}{\partial c^m} \right|_c = \int \rho_m^c \prod_{i=1}^m dV_i, \qquad (4)$$

where each $\int dV_i$ is an integration over the entire system volume. In other words, $\int \rho_m^c \prod_{i=1}^m (c_i - c) dV_i$ provides a measure of how $m$-site interactions modify $f^{ex}$ due to compositional inhomogeneity. The total free energy of the system is then given by

$$F[c] = \int \left[ \hat{f}(c) + \sum_m \frac{1}{m!} \int \rho_m^c \prod_{i=1}^m (c_i - c) dV_i \right] dV, \qquad (5)$$

where $\hat{f}(c) = f^{id}(c) + \hat{f}^{ex}(c)$.

Equation (5) is a key result of the present work. Equation (5), along with the constraints in eq. (4), is a key result of the present work and a powerful basis for unifying and assessing the validity of a broad range of existing free energetic descriptions. For example, as we show below, this construct provides a mathematically explicit interpretation of the locality assumption inherent in gradient expansion-based GL at any order. Consequently, we find that using eq. (5) as a starting point and then imposing the locality assumption leads to more generalized versions of commonly employed functionals. Moreover, the resulting functionals are characterized by physically transparent parameters that can be directly linked to experimental measurements and calculations. Although not a focus of the present work, other formalisms may also be derived as special cases of eq. (5). For example, as shown in Supplementary A, the Giacomin-Lebowitz model of phase separation[15], an example of a so-called 'non-local' GL theory that has recently gained much attention in the PDE community[32], can also be recovered as a special case of eq. (5). More generally, all non-local free energy formulations, such as those proposed in Refs.[33,34], must also satisfy eq. (5).

Finally, we note that it should be, in principle, possible to infer the complete convolutional kernels, $\rho_m^c$, directly from data (e.g., from compositional profiles around inter-phase boundaries[3]), without invoking the locality assumption. However, a complete prescription of the $\{\rho_m^c\}$ from compositional data would require a very high degree of measurement precision and we defer further discussion of this possibility to future work.

## Relation to the Ginzburg-Landau Formalism

In the GL formalism, the free energy density is assumed to be analytical with respect to gradient terms[8,21], i.e.,

$$F[c] = \int \left( \hat{f}(c) + \nabla c \cdot \boldsymbol{\kappa}(c) \cdot \nabla c + \cdots \right) dV, \quad (6)$$

where $\boldsymbol{\kappa}(c)$ is a symmetric matrix. The task at hand, therefore, is to determine the conditions under which eq. (5) may be stated in the form of eq. (6). Taylor expansion of each site composition, $c_i$, with respect to a reference composition $c = c(\mathbf{r})$, where $\mathbf{r}$ is the location at which the integrand in eq. (5) is being evaluated gives (see Methods B)

$$F[c] = \int \left[ \hat{f}(c) + \sum_m \sum_J \chi_{m,J} \left( \frac{\partial c}{\partial r} \right)^J \right] dV, \quad (7a)$$

where the coefficients $\chi_{m,\mathbf{J}}$ are

$$\chi_{m,\mathbf{J}} = \frac{1}{m!}\left(\frac{1}{j}\right)^{\mathbf{J}} \int \rho_m^c (\Delta \mathbf{r})^{\mathbf{J}} \prod_{i=1}^{m} dV_i, \qquad (7b)$$

and $\mathbf{J} = \{j_{i,\alpha}\}$, $j_{i,\alpha} \geq 1$ is a multi-index that runs over all possible gradient terms where $i \in \{1, \ldots, m\}$ and $\alpha$ represents the contribution along the $\alpha$ coordinate. Because eq. (7) contains all possible combinations of $\left(\frac{\partial c}{\partial r}\right)^{\mathbf{J}}$ that are allowed by symmetry, it is formally equivalent to eq. (6). But it also lays bare a known limitation of the GL framework[24,33]—not all function moments of $\rho_m^c$ are well defined (see Methods B). In other words, assumption of a gradient expansion, or equivalently the transformation of eq. (5) into eq. (7), places strong constraints, often referred to as *locality*[35], on the class of $\rho_m^c$ that are allowable. Consequently, continuum models based on gradient expansions, such as those proposed in Refs. [31,36], only include a subset of the most general free energies that can be proposed. Perhaps equally importantly, the equivalence between eqs. (6) and (7) provides a pathway for determining how the parameters of GL free energies are related to function moments of $\rho_m^c$, enabling the interpretation of GL parameters in terms of interatomic potentials, coarse-grained interaction models, or experimental phase diagram data. This point is demonstrated in the following section for the specific case of the Cahn-Hilliard free energy.

## Square-Gradient Theories and the Cahn-Hilliard Equation

We now consider in detail the specific case of second-order gradient expansion (eq. 6), which is often referred to as a square-gradient/GL/CH free energy and is a common basis of continuum and phase-field modeling of critical phenomena[1], where $\boldsymbol{\kappa}$ is usually assumed to be constant (but not necessarily isotropic). The most general $\boldsymbol{\kappa}$ is given by eq. (7) up to $\sum j_{i,\alpha} = 2$ terms (see Methods C), i.e.,

$$\kappa_{\alpha,\beta} = \left[\chi_2(\alpha,\beta) - \delta_{\alpha\beta}\frac{\partial \chi_1(\alpha)}{\partial c}\right], \qquad (8)$$

where $\alpha$ and $\beta$ are direction indices and $\chi_1(\alpha) \equiv \int \rho_1^c (\Delta r_{1,\alpha})^2 dV_1$ and $\chi_2(\alpha,\beta) \equiv \int\int \rho_2^c (\Delta r_{1,\alpha} \Delta r_{2,\beta}) dV_1 dV_2$ are the second moments of $\rho_1^c$ and $\rho_2^c$, respectively. Equation (8) may be rewritten as

$$\boldsymbol{\kappa} \equiv -\boldsymbol{\sigma}^2 \frac{\partial^2 \hat{f}^{ex}}{\partial c^2}, \qquad (9)$$

where $\kappa$ and $\sigma$ are the matrix forms of $\kappa_{\alpha,\beta}$ and $\sigma_{\alpha,\beta}$, respectively, $\sigma_{\alpha,\beta}^2 = \int\int[\delta_{\alpha\beta}\delta(r_{1,\alpha} - r_{2,\beta}) - 1]\tilde{\rho}_2^c(\Delta r_{1,\alpha}\Delta r_{2,\beta})dV_1dV_2$, and $\tilde{\rho}_2^c \equiv \rho_2^c / \frac{\partial^2 \hat{f}^{ex}}{\partial c^2}$ (see Methods C).

Equation (9) is an illuminating result in several regards. First, it shows that the square-gradient GL formalism is *most generally* expressed in terms of the gradient of the excess chemical potential and only the second moments of the convolutional kernels, $\rho_1^c$ and $\rho_2^c$. In other words, the square-gradient approximation does not (and cannot) include a complete description of these kernels. Conversely, assuming that the conditions are met for the square-gradient picture to be valid, only the second moments need to be estimated from data. This is a significant simplification because $\sigma$ represents a physically interpretable quantity—an effective interaction range between sites—that is generally straightforward to estimate. Moreover, while $\sigma$ may be composition-dependent, in practice, it is likely to be only weakly so. Equation (9) also provides the necessary and sufficient conditions for the commonly-employed constant $\kappa$ approximation[21] in the Cahn-Hilliard picture. While Cahn and Hilliard proved in their original papers that the constant $\kappa$ approximation is valid for a regular solution, there has been no rigorous proof of how it arises as the limit of general microscopic models[15,32]. But perhaps more significantly, it also proves that the forms of $\kappa$ obtained previously for special cases of regular solution models[21,22], most notably by Liu et al[22], may be rigorously extended to any solution thermodynamics and at any length scale.

Based on the preceding arguments, we consider a physically grounded relaxation of the constant $\kappa$ picture by only assuming $\sigma^2$ to be constant. The corresponding diffusion equation is given by (see Methods D)

$$\frac{\partial c}{\partial t} = \nabla \cdot \left\{ D\nabla\beta \left[ \hat{\mu} + \frac{1}{2}\left( \nabla \cdot \sigma^2 \cdot \nabla\hat{\mu}^{ex} + \frac{\partial \hat{\mu}^{ex}}{\partial c} \nabla \cdot \sigma^2 \cdot \nabla c \right) \right] \right\}, \quad (10)$$

where $\hat{\mu} = \frac{\partial \hat{f}}{\partial c}$ and $\hat{\mu}^{ex} = \frac{\partial \hat{f}^{ex}}{\partial c}$. Equation (10) shows explicitly how the chemical potential is modified by both compositional and chemical potential gradients. We note that a generalized GL theory by Gurtin[36], which has received considerable attention, is consistent with this picture. Finally, the Cahn-Hilliard equation may be obtained from eq. (10) by assuming $\hat{\mu}^{ex} = -\eta_{CH}c + b$, where $\eta_{CH}$ and $b$ are constants, giving

$$\frac{\partial c}{\partial t} = \nabla \cdot \{D\nabla\beta[\hat{\mu} - \nabla \cdot \kappa \cdot \nabla c]\}, \quad (11)$$

where $\kappa = 2\eta_{CH}\sigma^2$.

**Model Binary System Near a Tricritical Point**

In this section, a simple analytical free energy model is used to demonstrate some of the implications of eq. (10). Consider the family of symmetric free energy functions

$$\hat{f} = \hat{f}^{id} + \hat{f}^{ex} = [c \log c + (1-c) \log(1-c)] + \left[\alpha c(1-c) - \epsilon \exp\left(-\frac{(c-0.5)^2}{\gamma}\right)\right], \quad (12)$$

in which the second term in the square brackets, modulated by the adjustable parameter $\epsilon$, reflects deviations from regular solution behavior and results in a tricritical point. Note that in eq. (12), the concentration is normalized to represent the atomic fraction. Below the tricritical point, the free energy model in eq. (12) exhibits two stable 'phases': $\xi_1$ and $\xi_3$, where $c(\xi_1) = 0.07$ and $c(\xi_3) = 0.93$. Past the tricritical point, a third stable phase ($\xi_2$) emerges at $c(\xi_2) = 0.5$; see Supplementary C. We consider two distinct situations: (1) spinodal decomposition in initially homogeneous, subcritical systems ($\epsilon < 0.03$), and (2) pattern evolution in various supercritical settings ($\epsilon > 0.03$). The other parameters are fixed at $\alpha = 3.0$ and $\gamma = 0.03$. Connection to a 'best-fit' Cahn-Hilliard model for each value of $\epsilon$ is made by finding the value of the (constant) gradient energy parameter, $\eta_{CH}$, that minimizes the difference between $\hat{f}^{ex}(c)$ in eq. (12) and that of a regular solution, i.e., $\eta_{CH} = \underset{\eta_{CH}}{\text{Argmin}} \int \left[\hat{f}^{ex}(c) - \eta_{CH} c(1-c)\right] dx$. All simulations are conducted using a finite difference scheme (central difference) and periodic boundary condition on a $100 \times 100$ square grid, with uniform grid spacing fixed at $l_G = 5a_0$, where $a_0$ is the underlying lengthscale. The diffusion coefficient, $D$, is set to be constant across all simulation conditions.

*Spinodal Decomposition*: Shown in Fig. 2 are three cases in which a noisy uniform initial compositional distribution ($c = 0.5 + N(0, 0.01)$) undergoes spinodal decomposition. For each combination of $\epsilon$ and $\eta_{CH}$, the top row corresponds to the Cahn-Hilliard (CH) model prediction with $\kappa = 2\eta_{CH} \sigma^2$, while the bottom row is the generalized square-gradient (GSG) prediction with $\sigma = 2a_0$. As expected, the CH and GSG models predict identical spinodal decomposition evolution for $\epsilon = 0$ (regular solution), Fig. 2(a). However, as the excess free energy becomes increasingly non-quadratic (i.e., increasing $|\epsilon|$), the onset of spinodal decomposition predicted by the CH model is slowed considerably relative to the GSG model, Fig. 2(b,c). Moreover, there is an apparent difference in the dominant wavelength, with the CH model exhibiting a slightly finer pattern.

These observations may be quantitatively predicted in the context of a linear stability analysis (see Methods E). Specifically, the analysis shows that the dominant Fourier modes for the CH and GSG models diverge from each other as $\lambda_{GSG}^{max} - \lambda_{CH}^{max} \propto \sqrt{r}$ where $r \equiv \frac{\partial^2 \hat{f}^{ex}}{\partial c^2} / \frac{\partial^2 \hat{f}}{\partial c^2} \to \infty$ as the critical point is approached. The corresponding difference in spinodal decomposition timescale, $\tau_{GSG} - \tau_{CH} \propto r^2$, also diverges as the tricritical point is approached, as seen in Fig. 2.

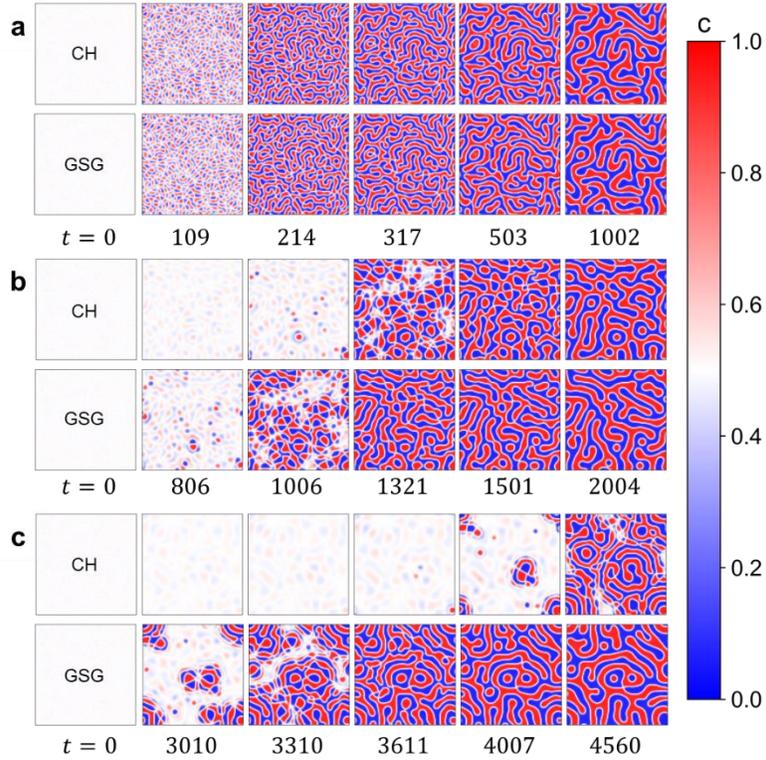

**Fig. 2 | Spinodal decomposition behavior in an initially homogeneous, subcritical mixture predicted by CH (top) and GSG (bottom) models for three mixture free energies**. **a** $\epsilon = 0$ and $\eta_{CH} = 3.0$, **b** $\epsilon = 0.022$ and $\eta_{CH} = 2.91$, and **c** $\epsilon = 0.026$ and $\eta_{CH} = 2.90$. In all cases **a**, **b**, **c**: $\sigma = 2a_0$ and time is scaled by the grid diffusion timescale, $\tau_D = (l_G)^2/D$.

*Pattern Evolution*: Motivated by the potential impact of thermal annealing on nanoscale devices, we next consider several supercritical systems with various initial compositional heterogeneities and several different parameter combinations. Shown in Fig. 3 is a situation in which a square region with area $100a_0^2$ and composition $c = 0.3$ is placed in the center of an otherwise homogeneous field at $c = 0.5$. In this set of simulations, we consider a supercritical system ($\epsilon = 0.075$ corresponding to $\eta_{CH} = 2.7$ for the best-fit CH model) with three different interaction ranges: (a) $\sigma = 2a_0$, (b) $2.3a_0$, and (c) $3.3a_0$. The GSG model results show a clear dependence on the interaction range parameter, $\sigma$. For a small interaction range, the gradient energy penalty is small, and the square region grows over time while maintaining a composition that corresponds to phase $\xi_1$. As the interaction distance is increased to $\sigma = 2.3a_0$, the growing patch exhibits a more rounded shape. Beyond this point, further increases to the interaction range destabilize the patch and lead to dissolution due to the gradient energy penalty becoming dominant. The CH model, however, predicts qualitatively different behavior. At the lowest interaction energy, the patch is observed to remain static over the simulation timescale. The patch does begin to grow and become more rounded as the interaction range increases, but the final trend towards dissolution is missed entirely in the CH picture.

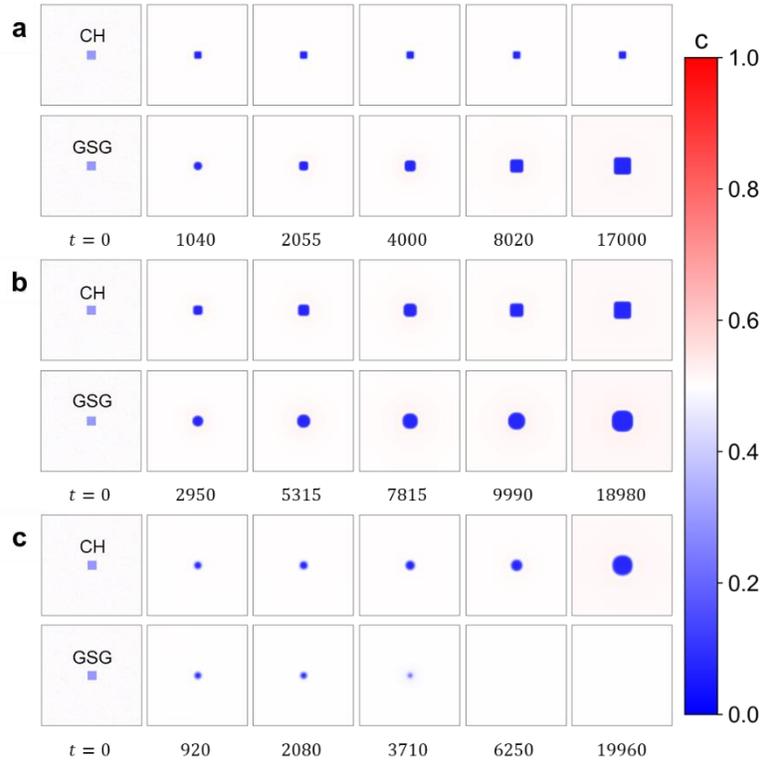

**Fig. 3 | Growth behavior of a square compositional heterogeneity predicted by CH (top) and GSG (bottom) models as a function of interaction range for a supercritical mixture**. **a** $\sigma = 2a_0$, **b** $\sigma = 2.3a_0$, **c** $\sigma = 3.3a_0$. For all cases **a**, **b**, **c**: $\epsilon = 0.075$, $\eta_{CH} = 2.70$, and time is scaled by the grid diffusion timescale, $\tau_D = (l_G)^2/D$.

Finally, we consider a strongly supercritical situation ($\epsilon = 0.1$, $\eta_{CH} = 2.6$ for the best-fit CH model) where the excess free energy curvature turns slightly positive near $c = 0.5$, and with an interaction range $\sigma = 3a_0$. As shown in Fig. 4, we investigate three initial configurations with different compositional heterogeneity geometries. The first two cases exhibit spinodal decompositions within the compositional heterogeneities in both GSG and CH models. However, in both instances, the spatiotemporal evolution predicted by the GSG model appears to be qualitatively more 'organized' and appears to produce higher symmetry configurations by the end of the simulations. This observation may be explained by the stronger gradient energy effects in the GSG description, which effectively delay a complete spinodal decomposition at early times. The delay allows the patterns in the GSG simulation to evolve more easily at earlier times, leading to the 'cleaner' final configurations. The last case, shown in Fig. 4(c), highlights yet another potential failure mode of the CH model. Here, two adjacent heterogeneities with compositions near $c(\xi_1) = 0.07$ and $c(\xi_3) = 0.93$ are, in principle, able to grow without altering the composition of the surroundings, which are initialized at $c(\xi_2) = 0.5$. The driving force for the growth of the heterogeneities is provided by the lower free energy of the $\xi_1$ and $\xi_3$ phases relative to $\xi_2$. This is indeed observed in the GSG model while the CH model predicts an essentially static situation.

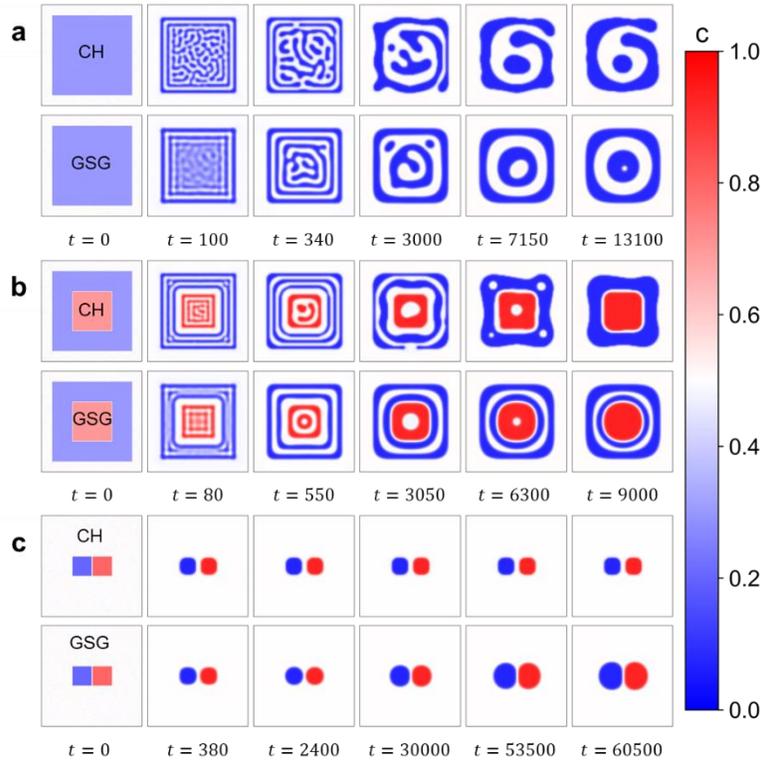

**Fig. 4 | Pattern evolution due to different initial compositional heterogeneities predicted by CH (top) and GSG (bottom) models in a supercritical mixture.** For all cases, **a**, **b**, **c**: $\epsilon = 0.1$, $\sigma = 3a_0$, and time is scaled by the grid diffusion timescale, $\tau_D = (l_G)^2/D$.

## Compound Semiconductor (SiGe) Interdiffusion

In this example, we consider interdiffusion in SiGe, a highly studied phenomenon with broad technological importance[37–42]. Our choice of this system is motivated by two characteristics. First, remarkably and somewhat uniquely, a large body of work[38,39,43] has established reasonable estimates for both the equilibrium thermodynamic and diffusion properties relevant to SiGe interdiffusion, enabling quantitative and predictive modeling. Second, we use the very simple phase behavior of the Si-Ge solid solution to demonstrate that gradient energy contributions may become significant in unexpected situations—most diffusion modeling in semiconductor systems assumes Fickian physics in which gradient energy contributions are neglected[37].

Literature data[37,38] is used to fully parametrize the GSG model, including the chemical potential function, the self-diffusivity, and the coherency strain contribution[21,37] (see Supplementary B). The only remaining parameter is the effective interaction range, $\sigma$, which we fix to be twice the lattice parameter, i.e., $2a_0 = 11.08$Å (see Supplementary C for additional results with $\sigma = a_0/2$ and $\sigma = a_0$). Interdiffusion was simulated in two QW-type configurations (denoted as 'well' and 'anti-well'), Fig. 5. The time evolution of the 'well' concentration profile predicted by the Fickian and GSG models is similar, showing a gradual spreading of the initial Gaussian configuration. On the other hand, the 'anti-well' configuration leads to qualitatively different evolution across the two models. Here, the Fickian model predicts slow diffusion in the center (low Ge fraction), which results in persistently sharp concentration peaks. The inclusion of the gradient term in the GSG model leads to much faster evolution and broadening. The differences between the two cases arise from the self-diffusivity's strong concentration dependence, which increases rapidly with increasing Ge fraction (see Supplementary B). In the 'well' configuration, diffusion is rapid in the center but becomes slower at the edges, effectively blocking the spread of Ge and reducing the impact of the gradient energy term. In contrast, the 'anti-well' configuration shifts the diffusion bottleneck to the center where the Ge fraction is lowest, and Fickian diffusion becomes very slow, resulting in the persistent peak. The addition of the gradient energy term, which enhances diffusion in the presence of large gradients, compensates for this effect in the GSG model. Given the ever-shrinking length scale (and potentially increasing sensitivity to interdiffusion-related degradation) of optoelectronic devices, we conclude that gradient energy effects may be necessary for modeling in these systems, even in the absence of apparent features such as phase separation.

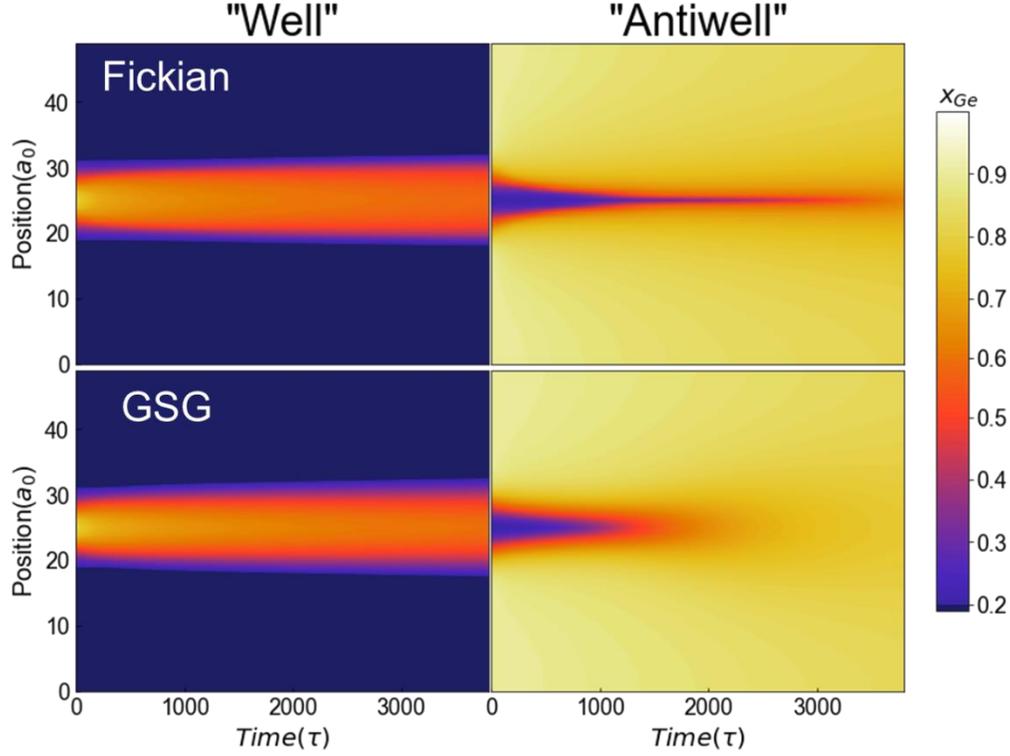

**Fig. 5 | One-dimensional interdiffusion as a function of time around a Gaussian well in SiGe.** Top row – Fickian, bottom row – GSG. 'Well' compositional profile (left) is given by $x_{Ge} = c_{Ge}/c_{tot} = 0.1 + 5.26 \cdot N(25, \sqrt{3})$, and 'anti-well' compositional profile (right) is $x_{Ge} = c_{Ge}/c_{tot} = 0.9 - 5.26 \cdot N(25, \sqrt{3})$. The quantity $c_{tot}$ is the concentration of atomic sites. Position is scaled by $a_0$. Time is scaled by the well variance diffusion timescale, $\tau = t/\tau_D$, with $\tau_D = 3/D_{max}$ and $D_{max} = D(x = 1)$. In all cases, $\sigma = 2a_0$.

## Discussion

The phenomenological nature of the GL free energy formalism has long been recognized as an important limitation of continuum dynamical models. As a result, it has often been difficult to make direct connections between key model parameters and microscopic physical properties, establish the bounds of model validity, or generalize models across material systems or even operating conditions. In this paper, we have presented a mathematically rigorous framework that leads to the most general hierarchy of free energy functionals in terms of a sequence of convolution kernels. These kernels are only weakly constrained by derivatives of the excess free energy of the system and therefore require additional inputs either from experimental measurements of compositional evolution or from microscopic (e.g., atomistic) simulations. Importantly, we demonstrate that the general hierarchy developed here can be explicitly matched to the gradient-expansion framework of the GL formalism. This matching provides precise mathematical insight into the nature of the approximations embodied within the GL construct while also demonstrating one possible pathway for approximating the convolutional kernels in the hierarchy developed here.

Looking ahead, several potential avenues for future study are apparent. Most obviously, the present hierarchy also may be used to analyze higher-order gradient expansions, such as the 4$^{th}$-order Swift-Hohenberg free energy[44], for which an analogous set of validity conditions may be obtained. The formalism developed here is also useful for establishing a rigorous connection to microscopic physics, most notably those represented by interatomic potential models[45–47]. One possible pathway for accomplishing this connection is to proceed via the classical DFT framework, where the equilibrium liquid-state direct correlation function may be used to infer the relevant convolution kernels. More broadly, the hierarchy developed here provides a formal mechanism for constructing a dictionary between thermodynamic properties of the system at equilibrium, e.g., interface shapes between different phases, and the free energy functional.

**References and Bibliography:**

- 1. Hohenberg, P. C. & Halperin, B. I. Theory of dynamic critical phenomena. *Rev. Mod. Phys.* **49**, 435–479 (1977).
- 2. Härtel, A. *et al.* Tension and stiffness of the hard sphere crystal-fluid interface. *Phys. Rev. Lett.* **108**, 226101 (2012).
- 3. Bhattacharyya, S., Sahara, R. & Ohno, K. A first-principles phase field method for quantitatively predicting multi-composition phase separation without thermodynamic empirical parameter. *Nat. Commun.* **10**, 3451 (2019).
- 4. Mianroodi, J. R. *et al.* Atomistic phase field chemomechanical modeling of dislocation-solute-precipitate interaction in Ni–Al–Co. *Acta Mater.* **175**, 250–261 (2019).
- 5. Vidyasagar, A., Krödel, S. & Kochmann, D. M. Microstructural patterns with tunable mechanical anisotropy obtained by simulating anisotropic spinodal decomposition. *Proc. Math. Phys. Eng. Sci.* **474**, 20180535 (2018).
- 6. Horsley, E. M., Lavrentovich, M. O. & Kamien, R. D. Aspects of nucleation on curved and flat surfaces. *J. Chem. Phys.* **148**, 234701 (2018).
- 7. Cao, D. *et al.* Lithium dendrite in all-solid-state batteries: Growth mechanisms, suppression strategies, and characterizations. *Matter* **3**, 57–94 (2020).
- 8. Chen, L.-Q. Phase-field models for microstructure evolution. *Annu. Rev. Mater. Res.* **32**, 113–140 (2002).

- 9. Bracha, D., Walls, M. T. & Brangwynne, C. P. Probing and engineering liquid-phase organelles. *Nat. Biotechnol.* **37**, 1435–1445 (2019).
- 10. Mao, S., Chakraverti-Wuerthwein, M. S., Gaudio, H. & Košmrlj, A. Designing the morphology of separated phases in multicomponent liquid mixtures. *Phys. Rev. Lett.* **125**, 218003 (2020).
- 11. di Santo, S., Villegas, P., Burioni, R. & Muñoz, M. A. Landau-Ginzburg theory of cortex dynamics: Scale-free avalanches emerge at the edge of synchronization. *Proc. Natl. Acad. Sci. U. S. A.* **115**, E1356–E1365 (2018).
- 12. Wio, H. S. *et al.* D-dimensional KPZ equation as a stochastic gradient flow in an evolving landscape: Interpretation and time evolution of its generating functional. *Front. Phys.* **4**, (2017).
- 13. Chmiela, S., Sauceda, H. E., Müller, K.-R. & Tkatchenko, A. Towards exact molecular dynamics simulations with machine-learned force fields. *Nat. Commun.* **9**, 3887 (2018).
- 14. te Vrugt, M., Löwen, H. & Wittkowski, R. Classical dynamical density functional theory: from fundamentals to applications. *Adv. Phys.* **69**, 121–247 (2020).
- 15. Giacomin, G. & Lebowitz, J. L. Phase Segregation in Dynamics in Particle Systems with Long Range Interactions. I. Macroscopic Limit. *J. Stat. Phys.* **87**, (1997).
- 16. Grmela, M. & Öttinger, H. C. Dynamics and thermodynamics of complex fluids. I. Development of a general formalism. *Phys. Rev. E Stat. Phys. Plasmas Fluids Relat. Interdiscip. Topics* **56**, 6620–6632 (1997).
- 17. Yang, L. *et al.* Direct numerical simulation of mass transfer and mixing in complex two-phase systems using a coupled volume of fluid and immersed boundary method. *Chemical Engineering Science: X* **5**, 100059 (2020).
- 18. Wittkowski, R. *et al.* Scalar φ4 field theory for active-particle phase separation. *Nat. Commun.* **5**, 4351 (2014).
- 19. Archer, A. J., Robbins, M. J., Thiele, U. & Knobloch, E. Solidification fronts in supercooled liquids: how rapid fronts can lead to disordered glassy solids. *Phys. Rev. E Stat. Nonlin. Soft Matter Phys.* **86**, 031603 (2012).
- 20. Dehghan, M., Abbaszadeh, M., Khodadadian, A. & Heitzinger, C. Galerkin proper orthogonal decomposition-reduced order method (POD-ROM) for solving generalized Swift-Hohenberg equation. *Int. J. Numer. Methods Heat Fluid Flow* **29**, 2642–2665 (2019).
- 21. Cahn, J. W. & Hilliard, J. E. Free energy of a nonuniform system. I. interfacial free energy. *J. Chem. Phys.* **28**, 258–267 (1958).
- 22. Liu, Z.-K., Ågren, J. & Suehiro, M. Thermodynamics of interfacial segregation in solute drag. *Mater. Sci. Eng. A Struct. Mater.* **247**, 222–228 (1998).

- 23. Asadi, E. & Asle Zaeem, M. A review of quantitative phase-field crystal modeling of solid–liquid structures. *JOM (1989)* **67**, 186–201 (2015).
- 24. Lutsko, J. F. First principles derivation of Ginzburg–Landau free energy models for crystalline systems. *Physica A* **366**, 229–242 (2006).
- 25. Archer, A. J., Ratliff, D. J., Rucklidge, A. M. & Subramanian, P. Deriving phase field crystal theory from dynamical density functional theory: Consequences of the approximations. *Phys. Rev. E.* **100**, 022140 (2019).
- 26. Wu, K.-A. & Voorhees, P. W. Phase field crystal simulations of nanocrystalline grain growth in two dimensions. *Acta Mater.* **60**, 407–419 (2012).
- 27. Athreya, B. P., Goldenfeld, N. & Dantzig, J. A. Renormalization-group theory for the phase-field crystal equation. *Phys. Rev. E Stat. Nonlin. Soft Matter Phys.* **74**, 011601 (2006).
- 28. Öttinger, H. C., Peletier, M. A. & Montefusco, A. A framework of nonequilibrium statistical mechanics. I. role and types of fluctuations. *J. non-equilib. thermodyn.* **46**, 1–13 (2021).
- 29. Rudraraju, S., Van der Ven, A. & Garikipati, K. Mechanochemical spinodal decomposition: a phenomenological theory of phase transformations in multi-component, crystalline solids. *Npj Comput. Mater.* **2**, (2016).
- 30. Parry, A. O. & Rascón, C. The Goldstone mode and resonances in the fluid interfacial region. *Nat. Phys.* **15**, 287–292 (2019).
- 31. Cherfils, L., Miranville, A. & Peng, S. Higher-order anisotropic models in phase separation. *Adv. Nonlin. Anal.* **8**, 278–302 (2017).
- 32. Davoli, E., Scarpa, L. & Trussardi, L. Nonlocal-to-local convergence of Cahn-Hilliard equations: Neumann boundary conditions and viscosity terms. *Arch. Ration. Mech. Anal.* **239**, 117–149 (2021).
- 33. Chen, X., Caginalp, G. & Esenturk, E. A phase field model with non-local and anisotropic potential. *Model. Simul. Mat. Sci. Eng.* **19**, 045006 (2011).
- 34. Gajewski, H. & Zacharias, K. On a nonlocal phase separation model. *J. Math. Anal. Appl.* **286**, 11–31 (2003).
- 35. Hohenberg, P. C. & Krekhov, A. P. An introduction to the Ginzburg–Landau theory of phase transitions and nonequilibrium patterns. *Phys. Rep.* **572**, 1–42 (2015).
- 36. Gurtin, M. E. Generalized Ginzburg-Landau and Cahn-Hilliard equations based on a microforce balance. *Physica D* **92**, 178–192 (1996).
- 37. Xia, G. (maggie). Interdiffusion in group IV semiconductor material systems: applications, research methods and discoveries. *Sci. Bull. (Beijing)* **64**, 1436–1455 (2019).
- 38. Prokes, S. M., Glembocki, O. J. & Godbey, D. J. Stress and its effect on the interdiffusion in $Si_{1-x}Ge_x/Si$ superlattices. *Appl. Phys. Lett.* **60**, 1087–1089 (1992).

- 39. Kube, R. *et al.* Composition dependence of Si and Ge diffusion in relaxed Si1−xGex alloys. *J. Appl. Phys.* **107**, 073520 (2010).
- 40. Aubertine, D. B. & McIntyre, P. C. Influence of Ge concentration and compressive biaxial stress on interdiffusion in Si-rich SiGe alloy heterostructures. *J. Appl. Phys.* **97**, 013531 (2005).
- 41. Kaiser, D., Han, S. M. & Sinno, T. Parametric analysis of mechanically driven compositional patterning in SiGe substrates. *J. Appl. Phys.* **121**, 065303 (2017).
- 42. David, T. *et al.* New strategies for producing defect free SiGe strained nanolayers. *Sci. Rep.* **8**, (2018).
- 43. Paskiewicz, D. M., Savage, D. E., Holt, M. V., Evans, P. G. & Lagally, M. G. Nanomembrane-based materials for Group IV semiconductor quantum electronics. *Sci. Rep.* **4**, 4218 (2014).
- 44. Swift, J. & Hohenberg, P. C. Hydrodynamic fluctuations at the convective instability. *Phys. Rev. A Gen. Phys.* **15**, 319–328 (1977).
- 45. Wang, J. *et al.* Machine learning of coarse-grained molecular dynamics force fields. *ACS Cent. Sci.* **5**, 755–767 (2019).
- 46. Wang, W. & Gómez-Bombarelli, R. Coarse-graining auto-encoders for molecular dynamics. *Npj Comput. Mater.* **5**, (2019).
- 47. Wen, M. & Tadmor, E. B. Uncertainty quantification in molecular simulations with dropout neural network potentials. *Npj Comput. Mater.* **6**, (2020).

**Acknowledgements:**

T.S. acknowledges support from the National Science Foundation through grant number DMR-1808065. S.M.H and G.B. acknowledge support from the National Science Foundation through grant number DMR-1809095.

**Author Contributions:**

A.B.L. and T.S. wrote the manuscript, with feedback from L.M., B.D.R., G.B., and S.M.H.

A.B.L. developed the mathematical framework, performed the simulations and analyses with feedback from T.S., L.M., B.D.R., G.B., and S.M.H. T.S. supervised the theory development.

**Competing Interests:**

The authors declare no competing interests.

## Methods A: Derivation of the Excess Free Energy Functional

Consider first a one-dimensional lattice/discretization of a domain where $f^{ex}$ at any site $r$ is denoted by $f^{ex}(\{c_n\})$, where $c_n \equiv c(r_n + r)$ is the composition at the $n^{\text{th}}$-neighbor site relative to site $r$, $n \in \mathbb{Z}$ is an index over all sites, and $c$ is the composition at site $r$. We denote $f^{ex}(\{c_n\})$ evaluated at some spatially uniform reference atomic fraction, $c$, as

$$\hat{f}^{ex}(c) \equiv f^{ex}(\{c_n\})|_{\{c_n = c\}}. \tag{A.1}$$

For a uniform perturbation in atomic fraction, $\delta$, Taylor expansion of the l.h.s. of eq. (A.1) gives

$$\hat{f}^{ex}(c + \delta) - \hat{f}^{ex}(c) = \sum_m \frac{1}{m!} \delta^m \left.\frac{\partial^m \hat{f}^{ex}}{\partial c^m}\right|_c, \tag{A.2}$$

where $m \in \mathbb{N}^+$. Next, Taylor expansion of the r.h.s. of eq. (A.1) gives

$$f^{ex}(\{c_n = c + \delta\}) - f^{ex}(\{c_n = c\}) = \sum_m \frac{1}{m!} \delta^m \left.\frac{\partial^m f^{ex}}{\partial c^m}\right|_{\{c_n = c\}}. \tag{A.3}$$

where

$$\frac{\partial}{\partial c} = \sum_n \frac{\partial}{\partial c_n}, \tag{A.4}$$

and

$$\frac{\partial^m}{\partial c^m} = \frac{\partial^{m-1}}{\partial c^{m-1}} \frac{\partial}{\partial c} = \sum_{\mathbf{m}} \left(\frac{\partial}{\partial c}\right)^{\mathbf{m}}, \tag{A.5}$$

where the multi-index notation, $\mathbf{m} = (n_1, n_2, \ldots, n_m)$, is introduced to denote

$$\left(\frac{\partial}{\partial c}\right)^{\mathbf{m}} = \left[\prod_{i=1}^m \frac{\partial}{\partial c_{n_i}}\right]. \tag{A.6}$$

Applying eq. (A.6) to eq. (A.3) gives

$$f^{ex}(\{c_n = c + \delta\}) - f^{ex}(\{c_n = c\}) = \sum_m \frac{1}{m!} \delta^m \sum_{\mathbf{m}} \left.\left(\frac{\partial}{\partial c}\right)^{\mathbf{m}} f^{ex}\right|_{\{c_n = c\}}$$

$$= \delta \sum_{n_1} \left.\frac{\partial f^{ex}}{\partial c_{n_1}}\right|_{\{c_n = c\}} + \frac{1}{2}\delta^2 \sum_{n_1} \sum_{n_2} \left.\frac{\partial^2 f^{ex}}{\partial c_{n_1} \partial c_{n_2}}\right|_{\{c_n = c\}} + \cdots. \tag{A.7}$$

Note that $m = \dim(\mathbf{m})$ so that $\Sigma_{\mathbf{m}}$ always refers to an $m$-dimensional sum. Equating Eqs. (A.3) and (A.7) then gives

$$\left.\frac{\partial \hat{f}^{ex}}{\partial c}\right|_c = \sum_n \left.\frac{\partial f^{ex}}{\partial c_n}\right|_{\{c_n=c\}}, \tag{A.8}$$

and

$$\left.\frac{\partial^m \hat{f}^{ex}}{\partial c^m}\right|_x = \sum_{\mathbf{m}} \rho_{\mathbf{m}}(c) \equiv \sum_{\mathbf{m}} \left.\left(\frac{\partial}{\partial c}\right)^{\mathbf{m}} f^{ex}\right|_{\{c_n=c\}}, \tag{A.9}$$

where we define $\rho_{\mathbf{m}} \equiv \left(\frac{\partial}{\partial c}\right)^{\mathbf{m}} f^{ex}\big|_{\{c_n=c\}}$ as the contribution of the $m$-site (contributions from varying the composition by $\delta$ at locations $n_1, n_2, \ldots, n_m$) to $\frac{\partial^m \hat{f}^{ex}}{\partial c^m}$.

The preceding analysis assumed a uniform compositional profile. In general, however, each $c_n$ is a different $\delta_n$ away from the reference composition $c$, i.e., $\{c_n = c + \delta_n\}$. A Taylor expansion of $f^{ex}$ in this general case gives

$$f^{ex}(\{c_n = c + \delta_n\}) = \hat{f}^{ex}(c) + \sum_m \frac{1}{m!} \sum_{\mathbf{m}} \delta^{\mathbf{m}} \rho_{\mathbf{m}}(c), \tag{A.10}$$

where

$$\delta^{\mathbf{m}} = \prod_{i=1}^m \delta_{n_i} = \prod_{i=1}^m (c_{n_i} - c). \tag{A.11}$$

Note that eq. (A.10), although derived for a one-dimensional domain, holds for any number of dimensions because a countable union of countable sets also is countable. In dimensions higher than one, it is convenient to rewrite the summation over $\mathbf{m}$ as summations over all $\{\mathbf{r}_1, \ldots, \mathbf{r}_m\}$ and straightforward to propose a continuum analog of eq. (A.11) as

$$f^{ex}([c], \mathbf{r}) = \hat{f}^{ex}(c) + \sum_m \frac{1}{m!} \int \rho_m^c \prod_{i=1}^m (c_i - c)\, dV_i, \tag{A.12}$$

where the sum over all $m$-sites is replaced by an integral over $\{V_i\} \equiv \{V_1, V_2, \ldots, V_m\}$, $dV_i = d\Delta r_{i,1} \ldots d\Delta r_{i,d}$ is an integral over the $d$ spatial dimensions of $\Delta \mathbf{r}_i$, where $\Delta \mathbf{r}_i$ is the variable whose integral replaces the sum over the $\mathbf{r}_i$ sites. In other words, $\rho_{\mathbf{m}}(c)$ and $\rho_m^c(c, \{\Delta \mathbf{r}_i\})$ may be interpreted as equivalent discrete and continuous convolutional kernels, respectively.

## Methods B: Connection to Ginzburg-Landau Type Theories

To establish a connection between eq. (5),

$$F[c] = \int \left[ \hat{f}(c) + \sum_m \frac{1}{m!} \int \rho_m^c \prod_{i=1}^m (c_i - c) \, dV_i \right] dV, \quad (B.1)$$

and the GL formalism,

$$F = \int \left( \hat{f}(c) + \nabla c \cdot \kappa(c) \cdot \nabla c + \cdots \right) dV, \quad (B.2)$$

note first that the composition on each site, $c_i$, can also be related to a reference site $\mathbf{r}$ with composition $c$, via an additional Taylor expansion, as

$$c_i - c = \sum_j \frac{1}{j!} [\Delta \mathbf{r}_i \cdot \nabla]^j c \big|_{\mathbf{r}}, \quad (B.3)$$

where $\Delta \mathbf{r}_i \equiv \mathbf{r}_i - \mathbf{r}$. Substituting eq. (B.3) into the integral terms on the R.H.S. of eq. (B.1) gives

$$\sum_m \frac{1}{m!} \int \rho_m^c \prod_{i=1}^m (c_i - c) \, dV_i = \sum_m \frac{1}{m!} \int \rho_m^c \prod_{i=1}^m \left( \sum_{j_i} \frac{1}{j_i!} (\Delta \mathbf{r}_i \cdot \nabla)^{j_i} c \big|_{\mathbf{r}} \right) dV_i$$

$$= \sum_m \frac{1}{m!} \int \rho_m^c \sum_{\mathbf{j}} \frac{1}{\mathbf{j}!} (\Delta \mathbf{r} \cdot \nabla c)^{\mathbf{j}} \bigg|_{\mathbf{r}} dV_i. \quad (B.4)$$

where $\mathbf{j} = (j_1, \ldots j_m)$ and

$$\frac{1}{\mathbf{j}!} (\Delta \mathbf{r} \cdot \nabla c)^{\mathbf{j}} \equiv \prod_{i=1}^m \left[ \frac{1}{j_i!} (\Delta \mathbf{r}_i \cdot \nabla)^{j_i} c \big|_{\mathbf{r}} \right]. \quad (B.5)$$

Consequently eq. (B.1) becomes

$$F = \int \left[ \hat{f}(c) + \sum_m \frac{1}{m!} \int \rho_m^c \sum_{\mathbf{j}} \frac{1}{\mathbf{j}!} (\Delta \mathbf{r} \cdot \nabla c)^{\mathbf{j}} \bigg|_{\mathbf{r}} \prod_{i=1}^m dV_i \right] dV. \quad (B.6)$$

In order to establish a connection between eq. (B.6) and the GL formalism in eq. (B.2), we note that the latter is written explicitly in terms of gradients of composition at various orders. Consequently, the terms in eq. (B.6) must be reordered according to

$$F = \int \left[ \hat{f}(c) + \sum_m \sum_j \frac{1}{j!} \frac{1}{m!} \int \rho_m^c (\Delta \mathbf{r} \cdot \nabla c)^j \Big|_{\mathbf{r}} \prod_{i=1}^m dV_i \right] dV. \tag{B.7}$$

This rearrangement is necessary to obtain distinct coefficients for each gradient term such as **κ** in eq. (6). Critically, such rearrangements are only guaranteed to converge for absolutely convergent series (Fubini's Theorem). Therefore, equating eqs. (B.2) and (B.7) implies that the representation of a GL free energy through gradient expansions requires this nontrivial assumption to hold. A similar concern was raised in an attempt to derive a GL free energy from classical DFT[24].

We illustrate the problem described above by considering the $m = 1$ case for eq. (B.7), i.e.,

$$F_1 = \int \left[ \hat{f}(c) + \sum_j \frac{1}{j!} \int \rho_1^c (\Delta r_{1,\alpha})^j \frac{\partial^j c}{\partial r_\alpha^j}\Big|_{\mathbf{r}} dV_1 \right] dV. \tag{B.8}$$

Since $\frac{\partial^j c}{\partial r_\alpha^j}\Big|_{\mathbf{r}}$ is independent of $\mathbf{r}_1$, eq. (C.8) can be written as

$$F_1 = \int \left[ \hat{f}(c) + \sum_j \frac{1}{j!} \int \rho_1^c (\Delta r_{1,\alpha})^j dV_1 \frac{\partial^j c}{\partial r_\alpha^j}\Big|_{\mathbf{r}} \right] dV. \tag{B.9}$$

However, eq. (B.9) is only valid if all $\int \rho_1^c (\Delta r_{1,\alpha})^j dV_1$ are bounded, or equivalently that all $j^{th}$ function moments of $\rho_1^c$ are finite. This necessarily fails if $\rho_1^c \propto \frac{1}{|\mathbf{r}|^k}$ asymptotically, i.e., decays algebraically, because $(\Delta r_{1,\alpha})^j \propto |\mathbf{r}|^j$, and the integral term in eq. (B.9) becomes

$$\int \rho_1^c (\Delta r_{1,\alpha})^j dV_1 \propto \int \frac{1}{|\mathbf{r}|^{k-j}} dV_1, \tag{B.10}$$

which will diverge for $k - j \leq (d - 1)$. Similar arguments hold for all $\rho_m^c$ and higher-order gradient terms can only be included if the $\rho_m^c$ have finite moments at corresponding orders. Note that for most stable distributions the variance is not well-defined, and for most Pareto distributions even the mean is not well-defined. Such distributions are of broad interest in physical and economic models based on Brownian motion, and diffusion on networks or general metric spaces may also readily have such dependencies in the form of heavy-tailed distributions.

For completeness, we also demonstrate explicitly how eq. (B.2) can be obtained from eq. (B.1) when the rearrangement from (B.6) to (B.7) is valid. To do so, recall that $\Delta \mathbf{r}_i = \{\Delta r_{i,\alpha}\}$, where $\alpha \in \{1, \dots, d\}$, and $d$ is the spatial dimension of the system. $\Delta \mathbf{r}_i \cdot \nabla$ can therefore be written as

$$\Delta \mathbf{r}_i \cdot \nabla = \sum_\alpha \Delta r_{i,\alpha} \frac{\partial}{\partial r_\alpha}. \tag{B.11}$$

Using eq. (B.5),

$$\frac{1}{\mathbf{j}!}(\Delta \mathbf{r}_i \cdot \nabla c)^{\mathbf{j}} = \prod_{i=1}^{m}\left[\frac{1}{j_i!}\left(\sum_\alpha \Delta r_{i,\alpha}\frac{\partial}{\partial r_\alpha}\right)^{j_i} c\right], \tag{B.12}$$

so that

$$\sum_{\mathbf{j}} (\Delta \mathbf{r}_i \cdot \nabla c)^{\mathbf{j}} = \sum_{\mathbf{J}} \left(\Delta r \frac{\partial c}{\partial r}\right)^{\mathbf{J}}, \tag{B.13}$$

where the multi-index notation $\mathbf{J} \equiv \{j_{i,\alpha}\}$ is introduced such that

$$\left(\frac{1}{\mathbf{j}}\Delta r \frac{\partial c}{\partial r}\right)^{\mathbf{J}} \equiv \prod_{i=1}^{m} \frac{1}{(\sum_\alpha j_{i,\alpha})!} \prod_{\alpha=1}^{d}\left[\left(\Delta r_{i,\alpha}\frac{\partial}{\partial r_\alpha}\right)^{j_{i,\alpha}} c\right]. \tag{B.14}$$

Substituting eq. (B.14) into eq. (B.7) then gives

$$F = \int\left[\hat{f}(c) + \sum_m \frac{1}{m!}\int \rho_m^c \sum_{\mathbf{J}}\left(\frac{1}{\mathbf{j}}\Delta r \frac{\partial c}{\partial r}\right)^{\mathbf{J}}\Bigg|_{\mathbf{r}} \prod_{i=1}^{m} dV_i\right]dV, \tag{B.15}$$

which may be written compactly as

$$F = \int\left[\hat{f}(c) + \sum_m \sum_{\mathbf{J}} \chi_{m,\mathbf{J}}\left(\frac{\partial c}{\partial r}\right)^{\mathbf{J}}\right]dV, \tag{B.16}$$

where the coefficients $c_{m,\mathbf{J}}$ are

$$\chi_{m,\mathbf{J}} = \frac{1}{m!}\left(\frac{1}{\mathbf{j}}\right)^{\mathbf{J}} \int \rho_m^c (\Delta r)^{\mathbf{J}} \prod_{i=1}^{m} dV_i, \tag{B.17}$$

and the integrals are the function moments of $\rho_m^c$. Note that because the free energy is a scalar, all odd function moments of $\rho_m^c$ must be zero. As such, eq (B.17) shows that the GL framework is equivalent to requiring that the function moments of $\rho_m^c$ do not diverge at all orders or a truncation of higher-order terms, which is equivalent to neglecting or setting higher-order moments of $\rho_m^c$ to be zero.

## Methods C: Interpretation of the Square-Gradient Coefficient

By inversion symmetry, eq. (B.7) can be written as

$$F \approx \int \left[ \hat{f}(c) + \frac{1}{2} \int \rho_1^c (\Delta \mathbf{r}_1 \cdot \boldsymbol{\nabla})^2 c \, dV_1 + \frac{1}{2} \int \int \rho_2^c (\Delta \mathbf{r}_1 \cdot \boldsymbol{\nabla} c) dV_1 (\Delta \mathbf{r}_2 \cdot \boldsymbol{\nabla} c) dV_2 \right] dV + \cdots, \quad (C.1)$$

or

$$F = \int \left[ \hat{f}(c) + \frac{1}{2} \left( \sum_\alpha \chi_1(\alpha) \frac{\partial^2 c}{\partial r_\alpha^2} + \sum_\alpha \sum_\beta \chi_2(\alpha, \beta) \frac{\partial c}{\partial r_\alpha} \frac{\partial c}{\partial r_\beta} \right) \right] dV + \cdots, \quad (C.2)$$

where

$$\chi_1(\alpha) \equiv \int \rho_1^c (\Delta r_{1,\alpha})^2 dV_1 \quad (C.3)$$

and

$$\chi_2(\alpha, \beta) \equiv \int \int \rho_2^c \Delta r_{1,\alpha} \Delta r_{2,\beta} dV_1 dV_2 \quad (C.4)$$

are the 2$^{nd}$-order function moments of $\rho_1^c$ and $\rho_2^c$ respectively. By the divergence theorem, eq. (C.2) can be rewritten as

$$F = \int \left\{ \hat{f}(c) + \frac{1}{2} \sum_\alpha \sum_\beta \left[ \chi_2(\alpha, \beta) - \delta_{\alpha\beta} \frac{\partial \chi_1(\alpha)}{\partial c} \right] \frac{\partial c}{\partial r_\alpha} \frac{\partial c}{\partial r_\beta} \right\} dV + \cdots. \quad (C.5)$$

Defining

$$\kappa_{\alpha,\beta} \equiv \left[ \chi_2(\alpha, \beta) - \delta_{\alpha\beta} \frac{\partial \chi_1(\alpha)}{\partial c} \right] \quad (C.6)$$

gives

$$F = \int \left\{ \hat{f}(c) + \frac{1}{2} \boldsymbol{\nabla} c \cdot \boldsymbol{\kappa} \cdot \boldsymbol{\nabla} c \right\} dV + \cdots, \quad (C.7)$$

recovering eq. (6) in the main text.

A more physically transparent expression for $\boldsymbol{\kappa}$ is derived below. Differentiating eq. (C.3) gives

$$\frac{\partial \chi_1(\alpha)}{\partial c} = \int \frac{\partial \rho_1^c}{\partial c} (\Delta r_{1,\alpha})^2 dV_1. \quad (C.8)$$

Using the recursive property of the convolutional kernels (eq. (4) in the main text)

$$\left.\frac{\partial^m \hat{f}^{ex}}{\partial c^m}\right|_c = \int \rho_m^c \prod_{i=1}^{m} dV_i, \tag{C.9}$$

eq. (C.3) can be rewritten as

$$\frac{\partial \chi_1(\alpha)}{\partial c} = \int \left( \int \rho_2^c dV_2 \right) (\Delta r_{1,\alpha})^2 dV_1. \tag{C.10}$$

Rearranging terms in eq. (C.10) gives

$$\frac{\partial \chi_1(\alpha)}{\partial c} = \int \int \delta(r_{1,\alpha} - r_{2,\beta}) \rho_2^c (\Delta r_{1,\alpha} \Delta r_{2,\beta}) dV_1 dV_2. \tag{C.11}$$

Substituting eq. (C.11) into eq. (C.6) gives

$$\kappa_{\alpha,\beta} = \int \int [1 - \delta_{\alpha\beta} \delta(r_{1,\alpha} - r_{2,\beta})] \rho_2^c (\Delta r_{1,\alpha} \Delta r_{2,\beta}) dV_1 dV_2. \tag{C.12}$$

Next, applying eq. (C.9) for $m = 2$, i.e.,

$$\frac{\partial^2 \hat{f}^{ex}}{\partial c^2} = \int \int \rho_2^c dV_1 dV_2, \tag{C.13}$$

and since $\frac{\partial^2 \hat{f}^{ex}}{\partial c^2}$ is independent of $r_1$ and $r_2$, we obtain

$$\int \int \tilde{\rho}_2^c dV_1 dV_2 \equiv \int \int \left(\frac{\partial^2 \hat{f}^{ex}}{\partial c^2}\right)^{-1} \rho_2^c dV_1 dV_2 = 1. \tag{C.14}$$

where $\tilde{\rho}_2^c \equiv \left[\left(\frac{\partial^2 \hat{f}^{ex}}{\partial c^2}\right)^{-1} \rho_2^c\right]$ is a convolution in the $V_1 \times V_2$ space. Using eq. (C.14), eq. (C.12) can now be written as

$$\kappa_{\alpha,\beta} = -\frac{\partial^2 \hat{f}^{ex}}{\partial c^2} \int \int [\delta_{\alpha\beta} \delta(r_{1,\alpha} - r_{2,\beta}) - 1] \tilde{\rho}_2^c (\Delta r_{1,\alpha} \Delta r_{2,\beta}) dV_1 dV_2. \tag{C.15}$$

Note that the negative sign and reordering in eq. (C.15) is introduced because the contribution from $\chi_1(\alpha)$ is usually larger than $\chi_2(\alpha, \beta)$ (in the original CH derivation[21], $\chi_2 = 0$). The integral term in eq. (C.15) is a linear combination of the 2nd moments of $\tilde{\rho}_2^c$ and has units of length squared, and we define it as

$$\sigma_{\alpha,\beta}^2 \equiv \int \int [\delta_{\alpha\beta} \delta(r_{1,\alpha} - r_{2,\beta}) - 1] \tilde{\rho}_2^c (\Delta r_{1,\alpha} \Delta r_{2,\beta}) dV_1 dV_2, \tag{C.16}$$

leading to the final result

$$\kappa_{\alpha,\beta} = -\sigma_{\alpha,\beta}^2 \frac{\partial^2 \hat{f}^{ex}}{\partial c^2}. \tag{C.17}$$

or in matrix form

$$\boldsymbol{\kappa} = -\boldsymbol{\sigma}^2 \frac{\partial^2 \hat{f}^{ex}}{\partial c^2}. \tag{C.18}$$

## Methods D: Derivation of Generalized Square-Gradient (GSG) Model

Consider the GL formalism

$$F = \int \left( \hat{f}(c) + \boldsymbol{\nabla} c \cdot \boldsymbol{\kappa}(c) \cdot \boldsymbol{\nabla} c + \cdots \right) dV, \tag{D.1}$$

with the gradient energy parameter derived in eq. (B.18) subject to constant $\boldsymbol{\sigma}$. The corresponding variational derivative is then given by

$$\frac{\delta F}{\delta c} \approx \hat{\mu} + \frac{1}{2} \frac{\partial^3 \hat{f}^{ex}}{\partial c^3} \boldsymbol{\nabla} c \cdot \boldsymbol{\sigma}^2 \cdot \boldsymbol{\nabla} c - \boldsymbol{\nabla} \cdot \left( \frac{\partial^2 \hat{f}^{ex}}{\partial c^2} \boldsymbol{\sigma}^2 \cdot \boldsymbol{\nabla} c \right)$$

$$= \hat{\mu} + \frac{1}{2} \frac{\partial^3 \hat{f}^{ex}}{\partial c^3} \boldsymbol{\nabla} c \cdot \boldsymbol{\sigma}^2 \cdot \boldsymbol{\nabla} c - \frac{\partial^3 \hat{f}^{ex}}{\partial c^3} \boldsymbol{\nabla} c \cdot (\boldsymbol{\sigma}^2 \cdot \boldsymbol{\nabla} c) - \frac{\partial^2 \hat{f}^{ex}}{\partial c^2} \boldsymbol{\nabla} \cdot (\boldsymbol{\sigma}^2 \cdot \boldsymbol{\nabla} c)$$

$$= \hat{\mu} - \frac{1}{2} \frac{\partial^3 \hat{f}^{ex}}{\partial c^3} \boldsymbol{\nabla} c \cdot \boldsymbol{\sigma}^2 \cdot \boldsymbol{\nabla} c - \frac{\partial^2 \hat{f}^{ex}}{\partial c^2} \boldsymbol{\nabla} \cdot (\boldsymbol{\sigma}^2 \cdot \boldsymbol{\nabla} c). \tag{D.2}$$

In the case of $\boldsymbol{\sigma}^2 = \sigma^2 \mathbf{I}$, eq. (D.2) reduces to

$$\frac{\delta F}{\delta c} \approx \hat{\mu} - \frac{1}{2} \frac{\partial^2 \hat{f}^{ex}}{\partial c^2} \sigma^2 (\boldsymbol{\nabla} c)^2 - \frac{\partial^2 \hat{f}^{ex}}{\partial c^2} \sigma^2 \boldsymbol{\nabla}^2 c, \tag{D.3}$$

or

$$\frac{\delta F}{\delta c} \approx \hat{\mu} - \sigma^2 \left[ \frac{1}{2} \frac{\partial^2 \hat{\mu}^{ex}}{\partial c^2} (\boldsymbol{\nabla} c)^2 + \frac{\partial \hat{\mu}^{ex}}{\partial c} \boldsymbol{\nabla}^2 c \right]. \tag{D.4}$$

Finally noting that

$$\boldsymbol{\nabla}^2 \hat{\mu}^{ex} = \boldsymbol{\nabla} \cdot \left( \frac{\partial \hat{\mu}^{ex}}{\partial c} \boldsymbol{\nabla} c \right) = \frac{\partial^2 \hat{\mu}^{ex}}{\partial c^2} (\boldsymbol{\nabla} c)^2 + \frac{\partial \hat{\mu}^{ex}}{\partial c} \boldsymbol{\nabla}^2 c. \tag{D.5}$$

allows us to rewrite eq. (D.4) as

$$\frac{\delta F}{\delta c} \approx \hat{\mu} - \frac{1}{2} \sigma^2 \left( \boldsymbol{\nabla}^2 \hat{\mu}^{ex} + \frac{\partial \hat{\mu}^{ex}}{\partial c} \boldsymbol{\nabla}^2 c \right). \tag{D.6}$$

Substituting eq. (D.6) into the diffusion equation finally gives

$$\frac{\partial c}{\partial t} = \nabla \cdot \left\{ D\nabla\beta \left[ \hat{\mu} + \frac{1}{2}\left( \nabla \cdot \sigma^2 \cdot \nabla \hat{\mu}^{ex} + \frac{\partial \hat{\mu}^{ex}}{\partial c} \nabla \cdot \sigma^2 \cdot \nabla c \right) \right] \right\}. \tag{D.7}$$

## Methods E: Linear Stability Analysis within the Square-Gradient Theory

Here we present linear stability analyses of CH and GSG free energy functionals,

$$F_{CH} \equiv \int \left[ \hat{f} + \frac{1}{2}\kappa(\nabla c)^2 \right] dV, \tag{E.1}$$

And

$$F_{GSG} \equiv \int \left[ \hat{f} - \sigma^2 \frac{\partial^2 \hat{f}^{ex}}{\partial c^2} (\nabla c)^2 \right] dV, \tag{E.2}$$

respectively, where both $\sigma^2$ and $\kappa$ are assumed to be constant. Note that eqs. (E.1) and (E.2) are equivalent when $\frac{\partial^2 \hat{f}^{ex}}{\partial c^2} = -\kappa/2\sigma^2$. Consider a system at an initial composition $c(\mathbf{r}) = c_0 + \epsilon \cos(\mathbf{q} \cdot \mathbf{r})$, where $\epsilon \ll 1$. Taylor expanding $\hat{f}$ about $x_0$ up to 2$^{nd}$-order gives

$$F_{CH} \approx \int \left[ \hat{f}(c_0) + \frac{\partial \hat{f}}{\partial c}(c - c_0) + \frac{1}{2}\frac{\partial^2 \hat{f}}{\partial c^2}(c - c_0)^2 + \frac{1}{2}\kappa(\nabla c)^2 \right] dV, \tag{E.3}$$

and,

$$\delta F_{CH} = F_{CH} - \int \hat{f}(c_0) dV \approx \int \left[ \frac{1}{2}\frac{\partial^2 \hat{f}}{\partial c^2}(\epsilon \cos(\mathbf{q} \cdot \mathbf{r}))^2 + \frac{1}{2}\kappa(\epsilon q^2 \sin(\mathbf{q} \cdot \mathbf{r}))^2 \right] dV, \tag{E.4}$$

where $q = |\mathbf{q}|$. For small perturbations the saddle-point approximation gives

$$\frac{\delta F_{CH}}{V}(q) \approx \frac{\epsilon^2 q^2}{4}\left[ \frac{\partial^2 \hat{f}}{\partial c^2} + \kappa q^2 \right] \delta x, \tag{E.5}$$

where all quantities are evaluated at $x_0$. The system is stable with respect to perturbations with wave number $q$ when $\frac{\delta F_{CH}}{V}(q) > 0$ and unstable when $\frac{\delta F_{CH}}{V}(q) < 0$. For spinodal decomposition, $\frac{\partial^2 \hat{f}}{\partial c^2} < 0$ and $\kappa > 0$. Therefore, the system is unstable with respect to perturbations $q < q_{CH}^c$, or equivalently for $\lambda > \lambda_{CH}^c$, where $\lambda = 2\pi/q$. Repeating the same analysis for the GSG model, the critical wavelengths are then given by

$$\lambda_{CH}^c = 2\pi \sqrt{-\kappa \left( \frac{\partial^2 \hat{f}}{\partial c^2} \right)^{-1}}, \tag{E.6}$$

and

$$\lambda_{GSG}^{c} = 2\pi\sigma\sqrt{2r}. \tag{E.7}$$

where $r \equiv \frac{\partial^2 \hat{f}^{ex}}{\partial c^2} / \frac{\partial^2 \hat{f}}{\partial c^2}$.

To obtain the fastest-growing Fourier modes among all the modes that do not decay, we consider the diffusion equation for each free energy (eqs. (E.1) and (E.2)). For $F_{CH}$,

$$\frac{\partial c}{\partial t} \approx D\nabla^2 \frac{\delta F}{\delta c} \approx D\left[\frac{\partial^2 \hat{f}}{\partial c^2}\nabla^2 c - \kappa\nabla^4 c\right], \tag{E.8}$$

where $D$ is assumed to be constant for small fluctuations. Applying a time-varying perturbation of the form $\delta x = \epsilon \cos(\mathbf{q}\cdot\mathbf{r})\exp(\omega t)$ into eq. (E.8) gives

$$\omega = -\frac{D}{4}\kappa q^2\left[\frac{1}{\kappa}\frac{\partial^2 \hat{f}}{\partial c^2} + q^2\right]. \tag{E.9}$$

Solving for the $q$ that maximizes $\omega$ gives the fastest growing Fourier mode wavelength as

$$\lambda_{CH}^{max} = 2\pi\sqrt{-\left(\frac{\partial^2 \hat{f}}{\partial c^2}\right)^{-1} 2\kappa} = 2\pi\sqrt{-\left(\frac{\partial^2 \hat{f}^{ex}}{\partial c^2}\right)^{-1} 2\kappa r}, \tag{E.10}$$

for the CH free energy and

$$\lambda_{GSG}^{max} = 4\pi\sigma\sqrt{r}. \tag{E.11}$$

The dominant Fourier modes for the CH and GSG models diverge from each other as the tricritical point is approached ($r \to \infty$) according to

$$\lambda_{GSG}^{max} - \lambda_{CH}^{max} = 2\pi\sqrt{r}\left(2\sigma - \sqrt{-2\kappa\left(\frac{\partial^2 \hat{f}^{ex}}{\partial c^2}\right)^{-1}}\right). \tag{E.12}$$

The corresponding growth rates, $\omega_{CH}^{max}$ and $\omega_{GSG}^{max}$, which scale as $q^4$ as the tricritical point is approached (see eq. (E.9)), tend to zero as

$$\omega_{GSG}^{max} - \omega_{CH}^{max} = -\frac{D}{8}\left[2\sigma^2 \frac{\partial^2 \hat{f}^{ex}}{\partial c^2}(q_{GSG}^{max})^4 - \kappa(q_{CH}^{max})^4\right] \propto r^{-4}. \tag{E.13}$$

Therefore, the spinodal decomposition timescales, $\tau_{CH}^{max} = 1/\omega_{CH}^{max}$ and $\tau_{GSG}^{max} = 1/\omega_{GSG}^{max}$, both tend towards infinity as

$$\tau_{GSG}^{max} - \tau_{CH}^{max} \propto \frac{1}{\omega_{GSG}^{max}} - \frac{1}{\omega_{CH}^{max}} \approx r^4. \tag{E.14}$$

## Supplementary A: Giacomin-Lebowitz Model of Phase Separation

It is interesting to contrast the present formulation to that proposed by Giacomin and Lebowitz in 1997[1] for an alternative model of phase separation that recovers the Cahn-Hilliard equation in the appropriate limits and which has gained much attention in the PDE community[2]. Due to the use of convolutions, the Giacomin-Lebowitz proposal is also called a non-local GL free energy theory, and our framework may be considered in that spirit as well.

Let $F^{GL}(c) \equiv f^{id}(c) + \hat{f}^{ex}(c)$, where the superscript GL here indicates Giacomin-Lebowitz and

$$\rho_m^c = \begin{cases} \frac{1}{2}[K^{GL}(\mathbf{r},\mathbf{r}_1)\delta(\mathbf{r}_1 - \mathbf{r}_2) + K^{GL}(\mathbf{r},\mathbf{r}_2)\delta(\mathbf{r}_1 - \mathbf{r}_2)] & \text{if } m = 2 \\ 0 & \text{otherwise} \end{cases}, \quad (A.1)$$

where $K^{GL}$ is a convolutional kernel that is a function only of pair distances and independent of composition. Substituting eq. (A.1) into eq. (5) in the main text gives

$$E^{GL} = \int \left[ F^{GL}(x) + \int K^{GL}(\mathbf{r},\mathbf{r}_1)(c_1 - c)^2 dV_1 \right] dV, \quad (A.2)$$

where $E^{GL}$ denotes the free energy functional proposed by Giacomin and Lebowitz. Note that the original GL proposal also included specific constraints on $F^{GL}$ and $K^{GL}$ with relation to truncated/finite interaction ranges and Kac potentials[1]. Comparing eq. (5) in the main text and eq. (A.2) shows several potential issues with the GL formulation. Firstly, the GL model only includes terms with $\rho_2^c$ and ignores $\rho_1^c$ terms. Secondly, $F^{GL}$ and $K^{GL}$ are treated as independent variables, which may be problematic if eq. (4) in the main text is not satisfied. Thirdly, since $K^{GL}$ is independent of composition, the GL formulation amounts to requiring that $\frac{\partial^2 \hat{f}^{ex}}{\partial x^2}\Big|_c = \frac{\partial^2 (F^{GL} - f^{id})}{\partial x^2}\Big|_c = const$ for all composition, which is not physically justified in most systems (except for the case of regular solutions, which was already proven by Cahn and Hilliard).

## Supplementary B: Parameters for SiGe Interdiffusion

The composition and temperature-dependent self-diffusivities for Si and Ge, $D_{Si}(x,T)$ and $D_{Ge}(x,T)$ in cm$^2$/s are obtained from Refs. 3–5

$$D_{Si} = \exp[6.489 + 4.964(1 - c) - 7.829(1 - c)^2] \times$$
$$\exp(-\beta[4.76(1 - c) + 3.32c + 1.54c(1 - c)]), \quad (B.1)$$

and

$$D_{Ge} = \exp[6.636 + 8.028c - 11.318c^2] \times$$
$$\exp(-\beta[3.83(1-c) + 3.13c + 1.63c(1-c)]). \tag{B.2}$$

The chemical potentials of Si and Ge are, in units of J/mol, (Ref. 6)

$$\hat{\mu}_{Si} = \ln(1-c) + \frac{1}{RT}(8787 - 1339c)c^2, \tag{B.3}$$

and

$$\hat{\mu}_{Ge} = \ln c + \frac{1}{RT}(8787 - 1339c)(1-c)^2. \tag{B.4}$$

In the lattice reference frame, the effective chemical potential is defined as

$$\hat{\mu} \equiv \hat{\mu}_{Ge} - \hat{\mu}_{Si}, \tag{B.5}$$

and total mobility/self-diffusivity is given by

$$D(c) = c_{Si}^2 D_{Si}(x) + c_{Ge}^2 D_{Ge}(x). \tag{B.6}$$

The self-diffusivity and effective chemical potential are plotted as a function of composition below.

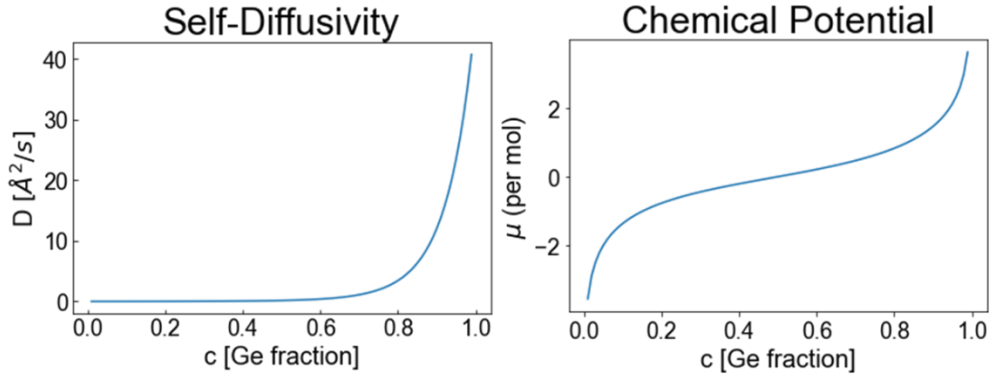

**Fig. B.1 | Self-diffusivity and chemical potential plots of of Si$_{1-x}$Ge$_x$ as functions of Ge atomic fraction. a** The effective self-diffusivity increases monotonically and nonlinearly with the Ge atomic fraction. **b** The monotonically increasing chemical potential implies the free energy is a convex function of Ge atomic fraction.

Since quantum wells are usually grown under unrelaxed conditions, the free energy is modified by an additional biaxial strain contribution. In this instance, we take the substrate composition $x_0$ to be the initial average composition outside the quantum well, and approximate it in the standard approach[7] for an isotropic solid (or equivalently, biaxial strain due to growth along the Si(100) direction, denoted as $z$) as

$$F^{el} = \int \eta^2 V_m \frac{E}{1-\nu}(c-c_0)^2 dz, \tag{B.7}$$

where $\eta$ is the percent lattice mismatch between Si and Ge, $V_m$ is the molar volume, $E$ is the Young's modulus, and $\nu$ is the Poisson ratio, with values obtained from Ref. 8:

$$E = (130.2 - 28.1c)\ GPa \tag{B.8}$$

$$\nu = (0.278 - 0.005c) \tag{B.9}$$

$$V_m = (12.06(1-c) + 13.63c)\ cm^3/mol. \tag{B.10}$$

Note that the strain contribution to the free energy is accounted for by simply adding eq. (B.7) to $F$ to obtain a total free energy of

$$F^{tot} = F + F^{el}, \tag{B.11}$$

where $F = F^{id} + F^{ex}$ is the free energy described in the main text.

## Supplementary C: Miscellaneous

Free energy diagram as a function of $\epsilon$ is included below.

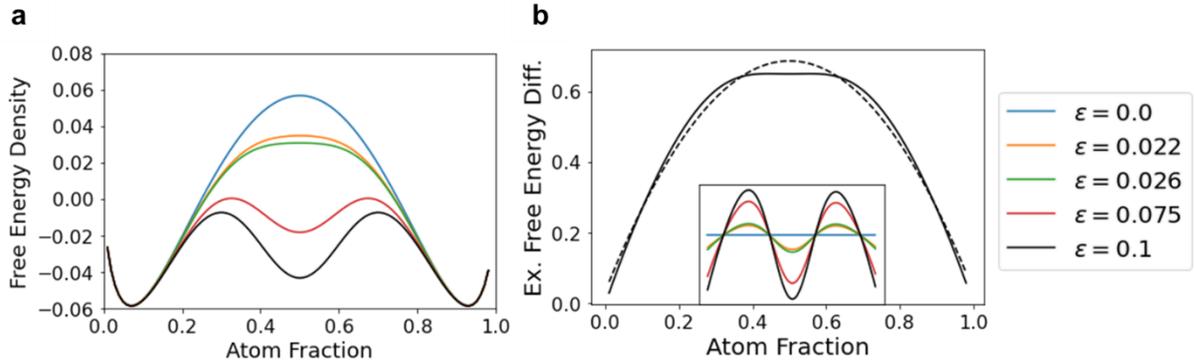

**Fig. C.1 | Free energy density plots for model in eq. (12). a** Free energy density plot as a function of composition and $\epsilon$. For $\epsilon \geq 0.75$, three minima develop. **b** Excess free energy density and its quadratic approximation (dotted line) for $\epsilon = 0.1$. The excess free energy differences are plooted as an inset figure, where the maximum amplitude in deviation is $< 0.02$.

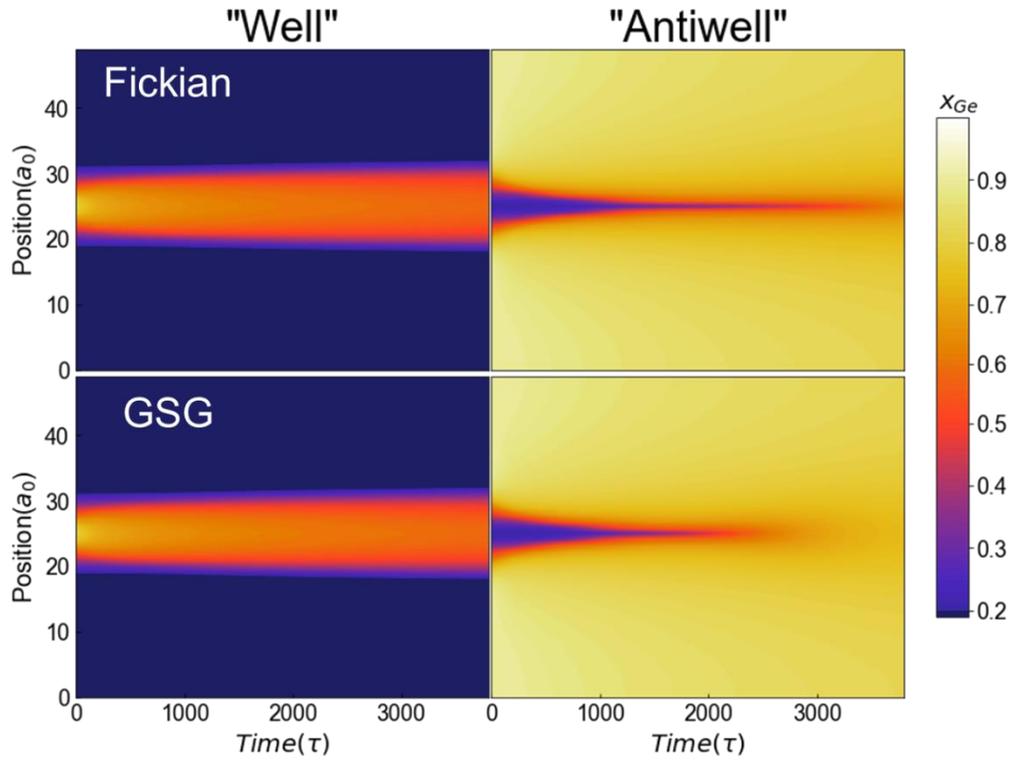

**Fig. C.2 | Comparison of SiGe interdiffusion prediction of Fickian and GSG models for $\sigma = a_0/2$.**

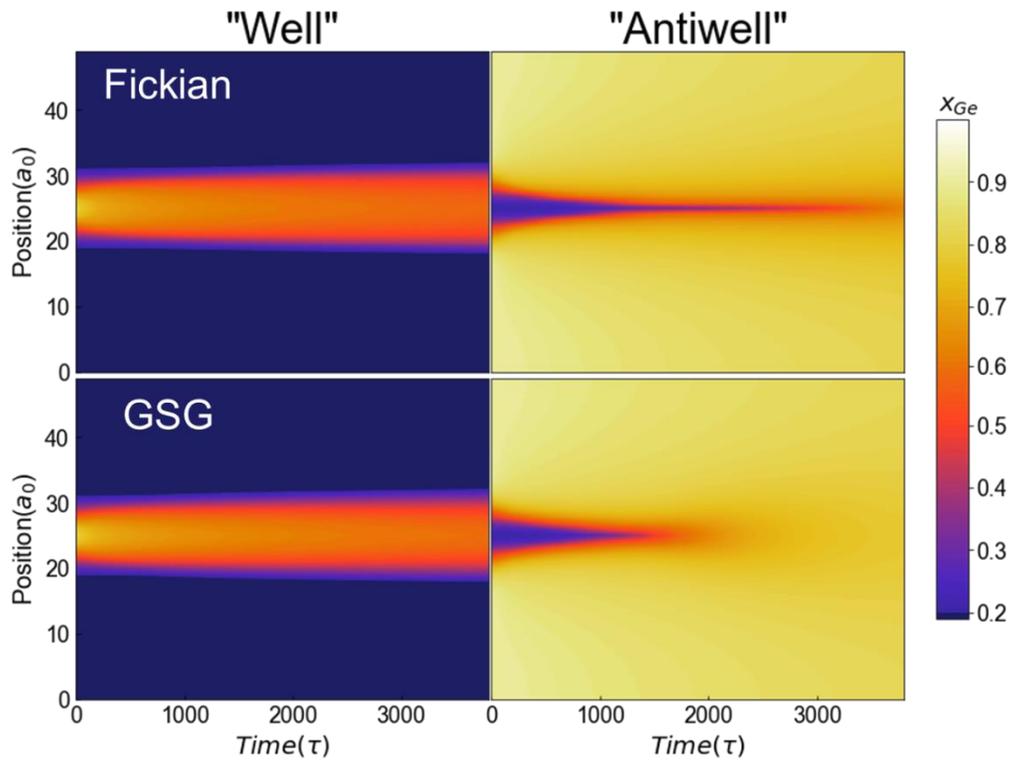

**Fig. C.3 | Comparison of SiGe interdiffusion prediction of Fickian and GSG models for $\sigma = a_0$.**

**References and Bibliography:**

- 1. Giacomin, G. & Lebowitz, J. L. Phase Segregation in Dynamics in Particle Systems with Long Range Interactions. I. Macroscopic Limit. *J. Stat. Phys.* **87**, (1997).
- 2. Davoli, E., Scarpa, L. & Trussardi, L. Nonlocal-to-local convergence of Cahn-Hilliard equations: Neumann boundary conditions and viscosity terms. *Arch. Ration. Mech. Anal.* **239**, 117–149 (2021).
- 3. Prokes, S. M., Glembocki, O. J. & Godbey, D. J. Stress and its effect on the interdiffusion in $Si_{1-x}Ge_x$/Si superlattices. *Appl. Phys. Lett.* **60**, 1087–1089 (1992).
- 4. Kube, R. *et al.* Simultaneous diffusion of Si and Ge in isotopically controlled heterostructures. *Mater. Sci. Semicond. Process.* **11**, 378–383 (2008).
- 5. Kube, R. *et al.* Composition dependence of Si and Ge diffusion in relaxed $Si_{1-x}Ge_x$ alloys. *J. Appl. Phys.* **107**, 073520 (2010).
- 6. Xia, G. (maggie). Interdiffusion in group IV semiconductor material systems: applications, research methods and discoveries. *Sci. Bull. (Beijing)* **64**, 1436–1455 (2019).
- 7. Cahn, J. W. & Hilliard, J. E. Free energy of a nonuniform system. I. interfacial free energy. *J. Chem. Phys.* **28**, 258–267 (1958).
- 8. Wortman, J. J. & Evans, R. A. Young's modulus, shear modulus, and Poisson's ratio in silicon and germanium. *J. Appl. Phys.* **36**, 153–156 (1965).